\documentclass[preprint,12pt,sort&compress]{elsarticle}
\usepackage{graphicx}
\usepackage{amssymb} 
\usepackage{amsbsy}  

\journal{Journal of Magnetism and Magnetic Materials}

\begin{document}

\begin{frontmatter}

\title{Training and recovery behaviours of exchange bias in FeNi/Cu/Co/FeMn spin valves at high field sweep rates}

\author[NTNU,LZU]{D. Z. Yang}
\author[NTNU]{A. Kapelrud}
\author[NTNU]{M. Saxegaard}
\author[NTNU]{E. Wahlstr\"{o}m}
\ead{erik.wahlstrom@ntnu.no}

\address[NTNU]{Institutt for fysikk, NTNU, NO-7491 Trondheim, Norway}
\address[LZU]{The Key Laboratory for Magnetism and Magnetic Materials
of Ministry of Education, Lanzhou University, Lanzhou 730000, China}

\begin{abstract}
\indent Training and recovery of exchange bias in FeNi/Cu/Co/FeMn
spin valves have been studied by magnetoresistance curves with field
sweep rates from 1000 to 4800 Oe/s. It is found that training and
recovery of exchange field are proportional to the logarithm of the
training cycles and recovery time, respectively. These behaviours
are explained within the model based on thermal activation. For the
field sweep rates of 1000, 2000 and 4000 Oe/s, the relaxation time
of antiferromagnet spins are 61.4, 27.6, and 11.5 in the unit of
$ms$ respectively, much shorter than the long relaxation time
($\sim10^{2} s$) in conventional magnetometry measurements.
\end{abstract}


\end{frontmatter}
\section{Introduction}
\indent The exchange bias (EB) effect in ferromagnetic/
antiferromagnetic systems have been intensely studied in the last
decade because of their physical complexity and important
applications \cite{Nogues1999,Berkowitz1999}. The technological
importance lies in the pinning effect of the antiferromagnet (AFM)
layers in which the hysteresis loop of the ferromagnet (FM) can be
shifted away from the origin point by the amount of the exchange
field ($H_\mathrm{E}$), and is usually accompanied with an enhanced
coercivity ($H_\mathrm{C}$). Changes of $H_\mathrm{E}$ and
$H_\mathrm{C}$ are accordingly directly related to the spin
configuration of the AFM layer through the exchange coupling
\cite{Qiu2008}. Among the variety of effects related to the EB
phenomenon, the training effect is an important effect that reflects
the AFM spin dynamic process during repeated hysteresis loops. It is
ascribed to that the spin structure of the AFM layer deviates from
its equilibrium configuration and approaches another equilibrium
triggered by subsequent reversals of the FM magnetization. Nowadays,
studies of AFM spin dynamic behaviours with training effect in both
experiments and theories have been widely reported
\cite{Dho2006,Heijden1998,Pina2004,Leighton2001,Hung2000,Carey2001,Chan2008,Hughes2001,Xi2007,Biternas2010}.
Because most of studies are limited to long timescales ($>1s$), by
the usually quite long measurement time in magnetometry approaches,
the relaxation time of AFM spin are usually reported in second
timescale ($\sim10^{2}-10^{4} s$)\cite{Dho2006, Heijden1998,
Pina2004}. In contrast, at shorter measurement timescales the
relaxation time of exchange bias system was demonstrated to cover a
wide range ( $\sim10^{-8}-10^{11} s$)
\cite{Garcia2002,Camarero2001,Goodman2001,Sahoo2007}, which  has
been ascribed to the magnetization reversal mechanism of FM layer
\cite{Raquet1995}. Hence, the report \emph{only} on AFM spin dynamic
behaviour in the millisecond timescale is still sparse. In addition,
recently attempt frequencies up to $10^{12}$ Hz in AFM layer have
been reported \cite{Fernandez2010}, which indicated the much shorter
relaxation timescale of AFM spin than earlier anticipated.
Therefore, it is necessary and interesting to study the AFM spin
dynamic process at
short timescale (technologic importance $<1 s$).\\
\indent In this work, we have studied the EB training and recovery
behaviours at the \emph{millisecond timescale} based on the
electrical transport measurements in FeNi/Cu/Co/FeMn spin valves.
The experiments show that at high field sweep rates recovery time of
exchange field after training procedures is three orders of
magnitude shorter than the values observed by usual magnetometry
techniques, and the relaxation of magnetoresistance (MR) is
demonstrated in the millisecond timescale. These clearly indicate
that AFM spin dynamic behaviours can be studied and resolved down to
the millisecond timescale utilizing the ordinary resistance measurements. \\
\section{Experiment and Results}
\indent The spin valves of Si (001)/Cu (10 nm)/Fe$_{20}$Ni$_{80}$ (3
nm)/Cu (3 nm)/Co (3 nm)/FeMn (8 nm)/Ta (3 nm)  were prepared by a
magnetron sputtering system. The base pressure was 2$\times $10$^{ -
5}$ Pa and the Ar pressure was 0.3 Pa during the deposition. The 10
nm Cu buffer layer was used to stimulate the fcc (111) preferred
growth of the FeMn layer in order to enhance the EB.  A magnetic
field of 130 Oe was applied in the film plane during deposition to
induce the uniaxial anisotropy and thus the EB.  Magnetoresistance
(MR) measurements were performed to probe the switching behaviours
of the pinned layer for different subsequent hysteresis loops.  The
magnetic field was provided by home-built Helmholtz coils, and MR
was measured in real-time system with 2 $M/s$ sampling rate. To
study training and recovery of the EB, we first performed forty
consecutive MR measurements with a fixed field sweep rate to
characterize the training procedures. Then we stopped the magnetic
field sweep with an waiting time $t$. Finally ten consecutive MR
measurements with the same field sweep rate were measured in order
to observe and confirm the EB recovery. For each sweep rate, $t$
varied from 0.1 to 10 $s$. \\

\begin{figure}[tb]
\begin{center}
\resizebox*{5 in}{!}{\includegraphics*{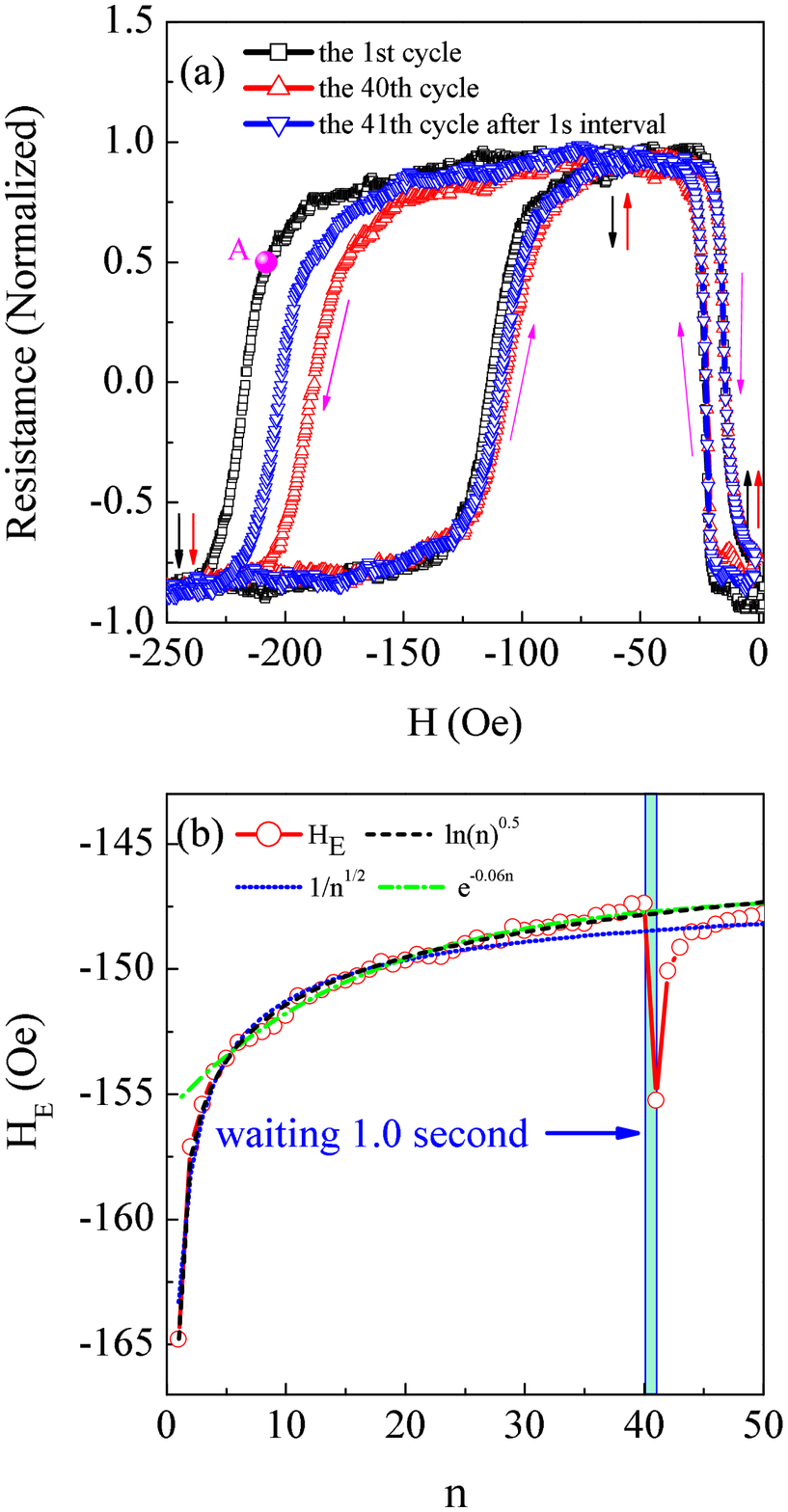}}
\end{center}
\caption{(a) The magnetoresistance curves used to map the training
effect for FeNi/Cu/Co/FeMn spin valve at the 1$^{st}$, 40$^{th}$ and
41$^{st}$ (after 1 s waiting time) cycles with the field sweep rate
of 4000 Oe/s. The resistance is dependent on corresponding
magnetization configurations of FeNi (black arrow) and Co (red
arrow). (b) The exchange field H$_E$ as a function of the number of
cycles $n$. The blue dot, green dash dot and black dash lines are
the fitted data with the $1/\sqrt{n}$ , $e^{-0.06n}$ and $ln(n)$,
respectively.} \label{Fig1}
\end{figure}

\begin{figure}[tb]
\begin{center}
\resizebox*{5 in}{!}{\includegraphics*{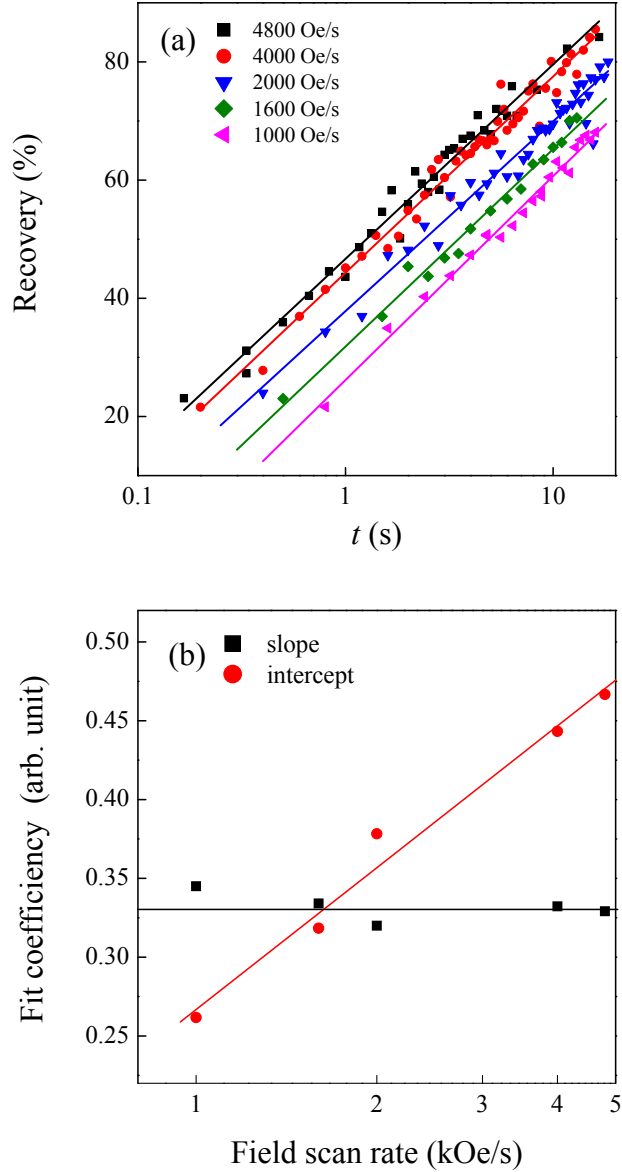}}
\end{center}
\caption{ (a) The recovery of $H_\mathrm{E}$ as a function of the
waiting time $t$ with different sweep rates. The solid lines display
the linear fits of the $ln(t)$. (b) The slope and the offset values
as a function of the field sweep rate. The solid lines are the
linear fits of the logarithm of the field sweep rate.} \label{Fig2}
\end{figure}

\begin{figure}[tb]
\begin{center}
\resizebox*{5 in}{!}{\includegraphics*{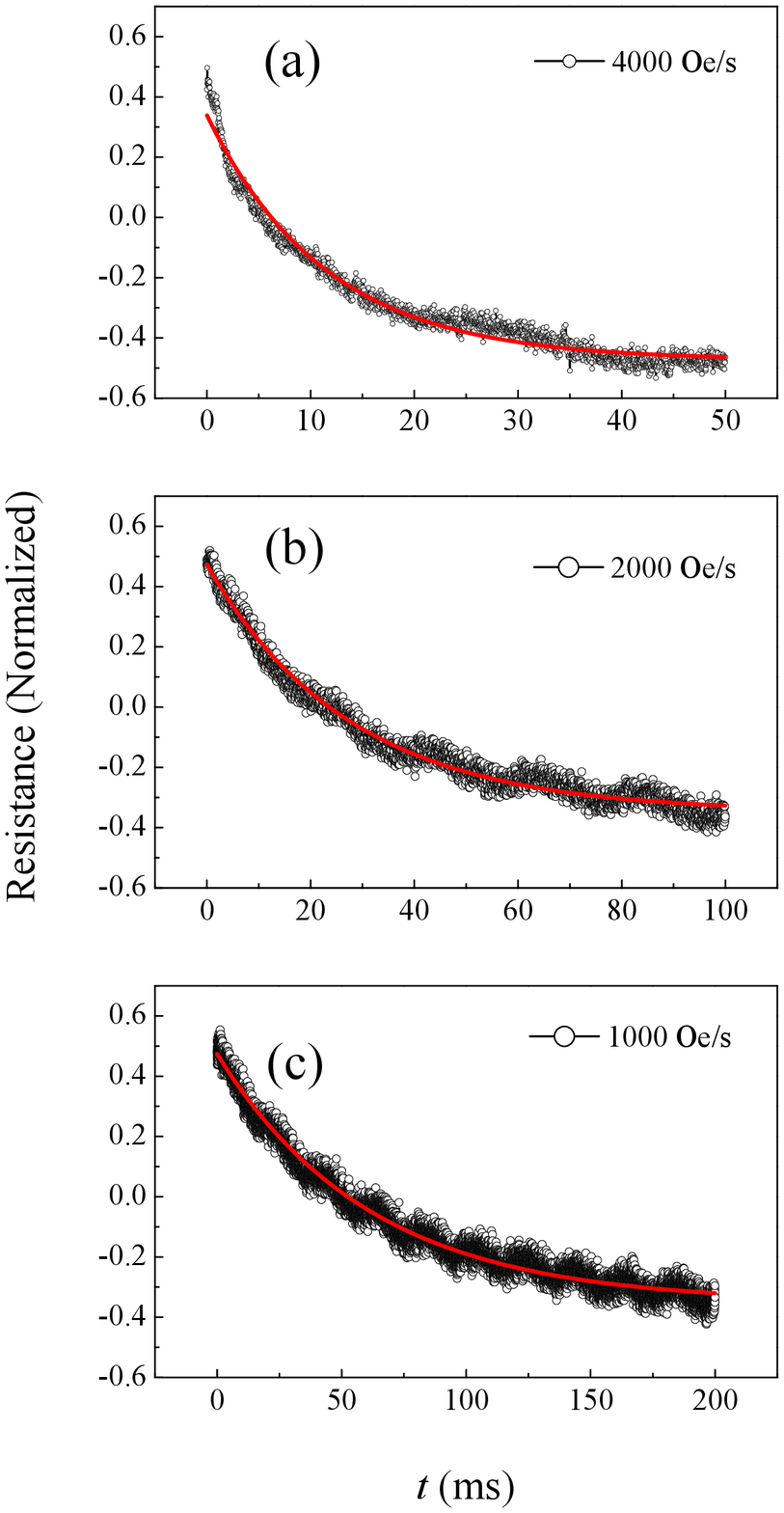}}
\end{center}
\caption{The time dependence of the resistance after the external
magnetic field is swept to -210 Oe (point A in Fig.~\ref{Fig1}(a))
from positive saturation field with different field sweep rates (a)
4000 Oe/s  (b) 2000 Oe/s (c) 1000 Oe/s. The solid lines are fits to
the first order exponential decay.} \label{Fig3}
\end{figure}

\indent The spin valves MR curves of the training and recovery
effects at the 1$^{st}$, 40$^{th}$ and 41$^{st}$ cycles with the
field sweep rate of 4000 Oe/s are displayed in Fig.~\ref{Fig1} (a).
At large negative field the Co and FeNi magnetizations are parallel
and pointing down. When the field is increased above the switch
field of the Co layer, about -110 Oe, the Co magnetization reverses
and resistance switches from low value (-1) to high value (+1). When
the field is further increased above the switch field of the FeNi
layer, about -15 Oe, its magnetization reverses, the two
magnetizations become parallel once more but this time pointing up,
and resistance switches to low value (-1). If the field is then
decreased, the two magnetizations will remain parallel until the
negative switch field of the FeNi layer is reached at -25 Oe, when
its magnetization reverses and resistance switches to high value.
When the field is further reduced and reverses the Co magnetization,
the two magnetizations align in parallel, and resistance changes to
its low value. For all MR curves the hysteresis loops of the Co
layer are shifted and fully separated from the hysteresis loops of
FeNi layer due to the FeMn pinning effect, therefore the MR curves
directly reflect the switching behaviours of the Co and the FeNi
layers in detail \cite{Ventura2008}. Comparing the hysteresis loops
of the Co layer in the 1$^{st}$, and 40$^{th}$ MR curves, the
switching field of the descent branch shifts more sharply than that
of the ascent one, demonstrating the asymmetric magnetization
reversal. However, after the magnetic field sweep is stopped for 1
$s$, a recovery is observed in the 41$^{st}$ MR curve. It contrasts
to the behaviour in the case of normally low field sweep rate, in
which substantial recovery was only observed after several hours of
waiting time ~\cite{Dho2006}. The $H_\mathrm{E}$ is plotted as
function of cycles $n$ in Fig.~\ref{Fig1}(b). The $H_\mathrm{E}$
gradually decreases with the cycle $n$, has an obvious resilience
after 1 $s$ waiting time and finally decreases. For the training
procedure, the $H_\mathrm{E}$ versus $n$ is fitted by a linear
functions of $1/\sqrt{n}$ , $e^{-0.06n}$ and $ln(n)$. It is found
that the logarithm function yields the best fit, except for initial
point
$n$=1~\cite{Hung2000,Chan2008}.\\
\indent To further study the recovery of the trained EB, we measured
the recovery rate $R$ as a function of $t$ at different field sweep
rates, where $R =[H_\mathrm E (41)-H_\mathrm E(40)]/[H_\mathrm
E(1)-H_\mathrm E(40)]\times 100(\% )$. Figure~\ref{Fig2}(a) shows
the dependence of $R$ on $t$ at different field sweep rates. The $R$
increases with the increasing  $t$ as a linear function of
\textit{log(t)}. More remarkably, for a fixed waiting time $t$, $R$
correspondingly increases with the increasing field sweep rate. This
logarithm behaviour is in a good agreement with the previous
experiments in NiFe/FeMn system, while the recovery rate is several
orders of magnitude faster than the value in the low field sweep
rate \cite{Dho2006}. The slope and the intercept as a function of
the field sweep rate are shown in Fig.~\ref{Fig2}(b). The slope
displays little change with different field sweep rates whereas the
intercept increases greatly as the field sweep rate increases in
approximate linear function of the logarithm of
the field sweep rate. \\
\indent  To investigate the dynamic behaviour of the EB with high
resolution, we observed the evolution of MR after setting the
magnetic field from the positive saturation field to -210 Oe (the
point A in Fig.~\ref{Fig1}(a)) near the switch field. As shown in
Fig.~\ref{Fig3}, MR initially decreases sharply and then gradually
reaches a constant. The small fluctuations in the curves are caused
by 50 Hz AC noise in the amplifying circuit. Remarkably, a crossover
of the normalized MR from positive to negative has been observed,
demonstrating the reversal of the magnetization of the Co layer. It
is possible to link the time dependence of MR with the magnetic
viscosity in the Co/FeMn bilayers \cite{Leighton2001}, in which the
magnetization of the pinned layer gradually reverses due to the
 thermally activated process in Co/FeMn bilayers.  Because
 the reversal process in the EB at the first cycle consists in the single domain wall motion \cite{Pina2004}, the
change of MR here is proportional to the amounts of the reversal
magnetization in the pinned layer. Shown as the solid line in
Fig.~\ref{Fig3},  the evolutions of MR are described well by a first
order exponential decay. From fitting the data, we extracted the
relaxation times $\tau $ are 11.5, 27.6, and 61.4 $ms$ for the field
sweep rate at 4000, 2000 and 1000 Oe/s respectively. This is again
in contrast to the long relaxation time ($\sim$800 $s$) in the
conventional approaches \cite{Pina2004}. One can also note the
relaxation time decreases with
the increasing field sweep rate. \\
\section{Discussion}
\indent The above results show that the recovery and relaxation of
the EB at high field sweep rates are faster than that earlier
observed \cite{Dho2006,Heijden1998,Pina2004,Leighton2001}. Below we
will interpret the
experimental results in conventional models for AFM and training effects.\\
\indent Firstly we consider the change and magnitude of the
relaxation time constants at different field sweep rates shown in
Fig.~\ref{Fig3}. The time constant for the relaxation can be
described by an ordinary Arrhenius law $\tau = v_\sigma ^{ - 1}\exp
(E_\sigma / k_B T)$, where $v_\sigma$ is the attempting frequency
and $E_\sigma=KV$ represents the AFM energy barrier, $K$ is the AFM
anisotropy and $V$ is the AFM grain volume. According to the AFM
grain volumes distribution, we can divide the $E_\sigma$ into three
different categories \cite{Chan2008}: $i)$ small $E_\sigma$ (small
grain size), which follows the FM magnetization at the timescale of
the experiment. $ii)$ medium energy $E_\sigma$ (medium grain size)
which will determine the EB dynamics at the timescale we
investigate. $iii)$ large $E_\sigma$ (large grain size), which is a
stable configuration over the timescale of the experiment. Assuming
 a typical uniaxial anisotropy constant of $1\times10^6$ erg/cm$^3$
and $v_\sigma$ to be $1\times10^{9}$ Hz, then the average grain size
of category (ii) is correspondingly about 9 nm extension, based on
the relaxation time in Fig.~\ref{Fig3}. Accordingly, that relaxation
time decreasing with the increasing field sweep rate, demonstrates
an apparent increase in attempt frequency $v_\sigma$.\\
\indent The EB recovery and relaxation at high field sweep rates can
still be explained well with the model based on thermal activation
\cite{Dho2006,Fulcomer1972}. As shown in Fig.~\ref{Fig2}(a), the
logarithm time recovery relationship indicates a thermally activated
reversal process involving the AFM spin configuration. To explain
our data, the activation energy spectrum model simply based on a
two-level system is adopted \cite{Fulcomer1972,Gibbs1983}. In our
case the two level system represents an individual AFM grain or
domain switching from a positive to a negative exchange energy with
respect to the FM layer. For the system with a wide energy barrier
distribution, $\Delta H_\mathrm E$ can be expressed in terms of the
AFM activation energy spectrum $q(E)$: $\Delta H_\mathrm E=q(E)k_B
Tln(v_\sigma t)$, which is taken from equation (1) in ref
\cite{Dho2006}. According to the equation, the slope observed for
all field rates in Fig.~\ref{Fig2}(b) is a constant due to the same
$q(E)$, while the intercept variation is mainly due to the different
activated AFM energy ranges and the time delay at the
different field sweep rates.\\
\indent Finally, for the training process $H_\mathrm {E}$ is
proportional to $ln(n)$ at high field sweep rates in
Fig.~\ref{Fig1}(b), which can be compared to the usual power law
($1/\sqrt{n}$) and the exponential ($e^{-\alpha n}$) relationships.
We model this through following Binek \textit{et
al}~\cite{Binek2004,Binek2006}. At beginning, the equilibrium AFM
interface magnetization is defined
$S^e_{AFM}=\lim_{n\rightarrow\infty}S_{AFM}(n)$. Each positive and
negative deviation $\delta S_n=S_{AFM}(n)-S^e_{AFM}$ of the AFM
interface magnetization from its equilibrium value will increase the
total free energy $F$ of the system by $\Delta F$. The relaxation of
the system towards equilibrium is determined by the
Landau-Khalatnikov (LK) equation\cite{Vizdrik2003}: $\xi
\dot{S}_{AFM} =-{\partial \Delta F}/{\partial S_{AFM}}$, where $\xi
$ is a phenomenological damping constant and  $\Delta F$ is the
function of $\delta S$. In Binek's model under the assumption
$\Delta F(\delta S) = \Delta F(-\delta S)$ , a series expansion of
$\Delta F$ up to the fourth order in $\delta S$ yields $\Delta F =
\frac{1}{2}a(\delta S)^2 + \frac{1}{4}b(\delta S)^4 + O(\delta
S)^6$. Evaluating the free energy expression with a leading term of
2$^{nd}$ and $4^{th}$
 order in $\delta S$ will result in the
$e^{-\alpha n}$ ~\cite{Binek2006} and $1/\sqrt{n}$ ~\cite{Binek2004}
evolution, respectively.

However, our understanding of the system is that we have a
nonvanishing odd order term. This is an effect of working at a time
scale where we also have substantial coupling at the FM/AFM
interface due to large grains that are too large to follow the
oscillating exchange coupling of the FM. Instead that portion of the
ensemble of grain will orient itself gradually according to the mean
coupling induced by the FM in a monotonic fashion. Accordingly we
also have to considered the expansion of $\Delta F$ also from  first
order of $\delta S$. We then assume $\delta S$: $\Delta F =
f(n)(\delta S)^1 + O(\delta S)^2$, where the $f(n)$ indicates that
the change in the AFM interface magnetization and $\delta S_n$ is
dependent on the training procedures $n$. By replacing $\dot{S}$
with $[S(n+1)-S(n)]/\tau$, with $\tau$ being the relevant
experimental time constant and the free energy expression of the
first order into the LK equation, we obtain an implicit sequence
equation: ${\xi}'(S(n+1)-S(n))=-f(n)$, where ${\xi}'={\xi}/{\tau}$.
The sum of this equation yields $H_\mathrm
E(n+1)\propto{S(n+1)}=-\sum{f(n)/{\xi}'}$.

Since we do not know the exact energy distribution of our system, we
make a first order approximation assuming a constant distribution of
areas, which leads to a linearly increasing activation energy with
time, then the approximate development $f(n)$ will follow a $1/n$
dependance. Using this approximation, we obtain a logarithmic
relaxation of the system that fits well with the observed
dependence. Note, that already at n=6 the approximation holds to
within an error of less than $4\%$. We note that the training
process $H_\mathrm {E}$ is proportional to $ln(n)$ has also been
reported at low field sweep rates~\cite{Hung2000}, where the
training speed is several orders
of magnitude slower than the values reported here.  \\
\indent In summary, for the AFM spins the relaxation time in the
millisecond timescale is demonstrated when the bilayers are exposed
to high field sweep rates.
 This behaviour can be well explained in terms of a time constrained
thermal activation. Our finding gives a new insight into the dynamic
behaviour of the AFM spins.\\
\section{Acknowledgments}
   We gratefully acknowledge helpful and fruitful
discussions with S. M. Zhou. The work was supported by the Norwegian
Research Council, Frinat project 171332. The author D. Z. Yang
acknowledges the funding supported by the National Natural Science
Foundation of China under Grand Nos. 11104122; the Fundamental
Research Funds for the Central Universities lzujbky-2011-51.




\begin{thebibliography}{30}
    \bibitem {Nogues1999}J. Nogues, I.K. Schuller, J. Magn. Magn. Mater. 192 (1999) 203.
    \bibitem {Berkowitz1999}A.E. Berkowitz, K. Takano, J. Magn. Magn. Mater. 200 (1999) 552.
    \bibitem {Qiu2008}X.P. Qiu, D.Z. Yang, S.M. Zhou, R. Chantrell, K. O'Grady, U. Nowak, J. Du, X.J. Bai, L. Sun, Phys. Rev. Lett. 101 (2008) 147207.
    \bibitem {Dho2006}J. Dho, C.W. Leung, M.G. Blamire, J. Appl. Phys. 99 (2006) 033910.
    \bibitem {Heijden1998}P.A.A. van der Heijden, T.F.M.M. Maas, W.J.M. de Jonge, J.C.S. Kools, F. Roozeboom, P.J. van der Zaag, Appl. Phys. Lett. 72 (1998) 492.
    \bibitem {Pina2004}E. Pina, C. Prados, A. Hernando, Phys. Rev. B 69 (2004) 052402.
    \bibitem {Leighton2001}C. Leighton, I.K. Schuller, Phys. Rev. B 63 (2001) 174419.
    \bibitem {Hung2000}C.Y. Hung, M. Mao, S. Funada, T. Schneider, L. Miloslavsky, M. Miller, C. Qian, H.C. Hong, J. Appl. Phys. 87 (2000) 4915.
    \bibitem {Carey2001}M.J. Carey, N. Smith, B.A. Gurney, J.R. Childress, T. Lin, J. Appl. Phys. 89 (2001) 6579.
    \bibitem {Chan2008}M.K. Chan, J.S. Parker, P.A. Crowell, C. Leighton, Phys. Rev. B 77 (2008) 014420.
    \bibitem {Hughes2001}T. Hughes, K. O'Grady, H. Laidler, R.W. Chantrell, J. Magn. Magn. Mater. 235 (2001) 329.
    \bibitem {Xi2007}H.W. Xi, S. Franzen, S.N. Mao, R.M. White,  Phys. Rev. B 75 (2007) 014434.
    \bibitem {Biternas2010}A.G. Biternas, R.W. Chantrell, U. Nowak, Phys. Rev. B 82 (2010) 134426.
    \bibitem {Raquet1995}B. Raquet, M.D. Ortega, M. Goiran, A.R. Fert, J.P. Redoules, R. Mamy, J.C. Ousset, A. Sdaq, A. Khmou, J. Magn. Magn. Mater. 150 (1995) L5.
    \bibitem {Garcia2002}F. Garcia, J. Moritz, F. Ernult, S. Auffret, B. Rodmacq, B. Dieny, J. Camarero, Y. Pennec, S. Pizzini, J. Vogel, IEEE. Trans. Mag. 38 (2002) 2730.
    \bibitem {Camarero2001}J. Camarero, Y. Pennec, J. Vogel, M. Bonfim, S. Pizzini, M. Cartier, F. Ernult, F. Fettar, B. Dieny, Phys. Rev. B 64 (2001) 172402.
    \bibitem {Goodman2001}A.M Goodman, K. O'Grady, H. Laidler, N.W. Owen, X. Portier, A.K. Petford-Long, F. Cebollada, IEEE. Trans. Mag. 37 (2001) 565.
    \bibitem {Sahoo2007}S. Sahoo, S. Plisetty, Ch. Binek, A. Berger, J. Appl. Phys. 101 (2007) 053902.
    \bibitem {Fernandez2010}G. Vallejo-Fernandez, N.P. Aley, J.N. Chapman, K. O'Grady, Appl. Phys. Lett. 97 (2010) 222505.
    \bibitem {Ventura2008}J. Ventura, J.P. Araujo, J.B. Sousa, A. Veloso, P.P. Freitas, Phys. Rev. B 77 (2008) 184404.
    \bibitem {Binek2004}Ch. Binek, Phys. Rev. B 70 (2004) 014421.
    \bibitem {Binek2006}Ch. Binek, S. Polisetty, X. He, A. Berger, Phys. Rev. Lett. 96 (2006) 067201.
    \bibitem {Fulcomer1972}E. Fulcomer, S.H. Charap, J. Appl. Phys. 43 (1972) 4190.
    \bibitem {Gibbs1983}M.R.J. Gibbs, J.E. Evetts, J.A. Leake, J. Mater. Sci. 18 (1983) 278.
    \bibitem {Vizdrik2003}G. Vizdrik, S. Ducharme, V.M. Fridkin, G. Yudin, Phys. Rev. B 68 (2003) 094113


\end{thebibliography}
\end{document}